\documentclass[a4paper]{article}
\usepackage{spconf,amsmath,graphicx}
\usepackage{hyperref}
\usepackage{import}
\usepackage{color}
\usepackage{subcaption}
\usepackage{multirow}
\usepackage{tabularx}
\usepackage{setspace}	
\usepackage{tikz}
\newcommand\QP{\mathit{QP}}

\def\VSPACE{-0.4cm}
\def\VSPACESECTION{-0.2cm}

\title{On Intra Video Coding and In-loop Filtering for Neural Object Detection Networks}
\name{Kristian Fischer, Christian Herglotz, and Andr\'e Kaup\thanks{The authors gratefully acknowledge that this work has been supported by the Deutsche Forschungsgemeinschaft (DFG) under contract number KA~926/10-1.}}
\address{Multimedia Communications and Signal Processing\\
	Friedrich-Alexander-Universit\"at Erlangen-N\"urnberg (FAU)\\
	Cauerstr. 7, 91058 Erlangen, Germany\\
	\{Kristian.Fischer, Christian.Herglotz, Andre.Kaup\}@fau.de}
\newcommand\copyrighttext{%
	\footnotesize \textcopyright 2020 IEEE. Personal use of this material is permitted.
	Permission from IEEE must be obtained for all other uses, in any current or future 
	media, including reprinting/republishing this material for advertising or promotional 
	purposes, creating new collective works, for resale or redistribution to servers or 
	lists, or reuse of any copyrighted component of this work in other works. 
	DOI: \href{https://doi.org/10.1109/ICIP40778.2020.9191023}{10.1109/ICIP40778.2020.9191023} }
\newcommand\copyrightnoticeOwn{%
	\begin{tikzpicture}[remember picture,overlay]
		\node[anchor=north,yshift=-10pt] at (current page.north) {\fbox{\parbox{\dimexpr\textwidth-\fboxsep-\fboxrule\relax}{\copyrighttext}}};
	\end{tikzpicture}%
}

\begin{document}
%
\maketitle
\copyrightnoticeOwn
\vspace{-4mm}
\begin{abstract}
Classical video coding for satisfying humans as the final user is a widely investigated field of studies for visual content, and common video codecs are all optimized for the human visual system (HVS). But are the  assumptions and optimizations also valid when the compressed video stream is analyzed by a machine? To answer this question, we compared the performance of two state-of-the-art neural detection networks when being fed with deteriorated input images coded with HEVC and VVC in an autonomous driving scenario using intra coding. Additionally, the impact of the three VVC in-loop filters when coding images for a neural network is examined. The results are compared using the mean average precision metric to evaluate the object detection performance for the compressed inputs. Throughout these tests, we found that the Bj\o ntegaard Delta Rate savings with respect to PSNR of 22.2~\% using VVC instead of HEVC cannot be reached when coding for object detection networks with only 13.6~\% in the best case. Besides, it is shown that disabling the VVC in-loop filters SAO and ALF results in bitrate savings of 6.4~\% compared to the standard VTM at the same mean average precision.
\end{abstract}
\vspace{-2mm}
\vspace{\VSPACESECTION}
\section{Introduction}
\label{sec:intro}

\begin{figure}[t]
	\centering
	\begingroup
	\footnotesize
	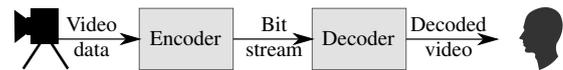
	\caption{Video coding for humans.}
	\label{fig:video coding humans}
	\endgroup
	\vspace{\VSPACE}
\end{figure}

\begin{figure}[t]
	\centering
	\begingroup
	\footnotesize
	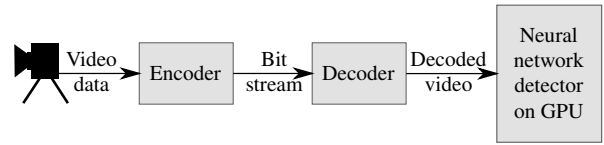
	\vspace{-2mm}
	\caption{Video coding for machines.}
	\label{fig:video coding machines}
	\endgroup
	\vspace{\VSPACE}
\end{figure}

Thanks to their outstanding performances, more and more applications are using neural networks that solve different tasks on multimedia data. However, these neural networks have the major drawback that they all need high-end graphic processing units (GPUs) which are first expensive and second not practical to battery driven devices like cars or unmanned aerial vehicles (UAVs). Thus, it is required to outsource the neural network execution to a remote server that is equipped with a GPU. However, the rate-constrained channels demand for compressing the multimedia data. Therefore, we investigate whether classical video coding for humans, as depicted in Fig.~\ref{fig:video coding humans}, also operates effectively when the decoded video is analyzed by a neural network detector as shown in Fig.~\ref{fig:video coding machines}.


%

Coding input data for neural networks can be attributed to the field of \textit{Video Coding for Machines (VCM)} that deals with machine to machine (M2M) communication. According to the Cisco Visual Networking Index \cite{cisco2019}, 51~\% of total devices and connections will be used for M2M communication systems in 2022, and the amount of M2M internet traffic will increase from 4~\% in 2017 to 7~\% in 2022 , so it is reasonable to put effort in coding M2M data. The significance of this topic is also underlined by the fact that MPEG introduced an ad hoc group on VCM in 2019 \cite{zhang2019}, which tackles several use cases like smart factory, video surveillance, and autonomous vehicles. In the scope of this paper we focus on the latter one.

One use case demanding for VCM is a collision avoidance system for an UAV as presented in \cite{lee2017}. There, the captured video data has to be sent to a cloud server that runs the neural detection network, since there is no place for the heavy and energy demanding GPU on the drone. 

Several works already exist in the field of compressing data for neural networks. In \cite{bagdanov2011}, the authors proposed a modification of the video coding standard H.264 that preserves important areas, while background pixels are compressed stronger.
Choi and Baji\'c proposed a new rate control method for H.265 that is adapted to object detection and spends more bits on areas that are vital for the used YOLO9000~\cite{redmon2017CVPR} object detection network~\cite{choi2018}.
Another work improves the object detection performance by creating saliency maps that helps the video encoder to focus on the relevant image parts \cite{galteri2018}.

More theoretical work has been done in \cite{dodge2016} where investigations were performed on the influence of blur, noise, and JPEG compression to neural classification networks. The same authors also found that humans perform better in solving classification tasks on noisy and blurred images than deep neural networks~\cite{dodge2017}. In \cite{ghosh2018}, a method for CNNs was designed to be more robust against noisy, blurred, or JPEG compressed images.

The investigations of this paper are twofold. First, we compare the performances of the High Efficiency Video Coding (HEVC) \cite{sullivan2012_HEVC} with its successor Versatile Video Coding~(VVC) \cite{chen2019vtm6} for intra coded frames in an autonomous driving scenario and two different region-based convolutional neural network (R-CNN) architectures. Second, we investigate the three VVC in-loop filters comparing their influence on the HVS and the R-CNNs for the given setup.

\vspace{\VSPACESECTION}
\section{Analytical Methods}
\label{sec:format}

\begin{figure}[t]
	\centering
	\resizebox{0.48\textwidth}{!}{
		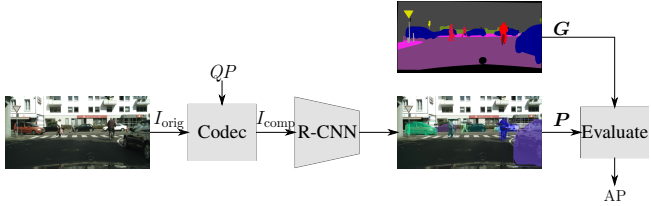
	}
	
	\caption{Signal flow investigating the codec performance for R-CNNs.}
	\label{fig:flow chart}
\end{figure}

In Fig. \ref{fig:flow chart}, the basic flow chart of our simulations is depicted. The input RGB image $I_\mathrm{orig}$, which should be analyzed, is first transformed to the YCbCr color format with 4:2:0 subsampling and then coded with either HEVC or VVC, controlling the quality with the quantization parameter~($\QP$). The resulting compressed image $I_\mathrm{comp}$ has to be back-converted to the RGB colorspace, before applying an R-CNN. Subsequently, the resulting predictions $\boldsymbol{P}$ from the network, which include the pixels belonging to the instance, the predicted class, and the certainty, are evaluated against the ground-truth~(GT) data~$\boldsymbol{G}$ with the average precision metric (AP).

\begin{table}[t]
	\caption{\textit{Cityscapes} object categories with the number of instances in validation set with `MC' and `BC' standing for motorcycle and bicycle, respectively.}
	\label{tab:cityscapes instances}
	\centering
	\begingroup
	\small
	\begin{tabularx}{\columnwidth}{c|c|c|c|c|c|c|c}

		Person & Rider & Car & Truck & Bus & Train & MC& BC\\
		\hline
		3399 & 544 & 4656 & 93 & 98 & 23 & 149 & 1169

	\end{tabularx}
	\endgroup
\vspace{\VSPACE}
\end{table}

\vspace{\VSPACESECTION}
\vspace{-1mm}
\subsection{Dataset}
\vspace{-1mm}
All simulations are conducted on the \textit{Cityscapes} dataset~\cite{cordts2016}, which includes automotive stereo data observing urban street scenes in Germany. From that dataset, the 500 uncompressed validation images of the left camera with a spatial resolution of $2048\times1024$ pixels are used. The GT data is labeled pixel-wise including multiple classes. For our purpose, we consider the eight classes listed in Table \ref{tab:cityscapes instances} with the number of instances occurring in the validation set.

For evaluating the object detection performance, the AP is used with the adaptations from the \textit{Cityscapes} challenge as described in \cite{cordts2016}. The AP value calculates the area under the precision-recall curve for several Intersection over Union~(IoU) thresholds and for each object category. Thereby,  $\mathrm{AP}_{50}$ is calculated for all predicted instances having an IoU with the GT instances above a threshold of 50~\%, whereas AP is averaged over ten IoU thresholds from 50~\% to 95~\% in steps of 5~\%. For the mean AP~(mAP) metric, the AP values of all object categories are averaged. Two modifications are made compared to the original \textit{Cityscapes} evaluation code for instance-level semantic labeling from \cite{cordts2017cityscapesscripts}. For Faster R-CNN, the GT pixel-wise annotations are converted to bounding boxes for a comparable calculation of IoU values. Second, the mAP is calculated with a weighted average based on the number of instances of each class, to avoid that classes with a low instance count (e.g. train) contribute as much to the mAP and therewith to the coding performance evaluation as classes with many instances (e.g. car).

\vspace{\VSPACESECTION}
\subsection{Investigated Codecs}

In the scope of this paper, we used the HEVC test model~(HM) software \cite{hm_software} in version 16.2 and the VVC test model (VTM) in version 6.0 for VVC \cite{chen2019vtm6}. We relinquish using traditional image codecs like JPEG, because they were shown to perform worse than HEVC on images \cite{nguyen2012}.

Since the \textit{Cityscapes} dataset only contains single images, the video codecs HEVC and its successor VVC are tested in the all-intra configuration.

\vspace{\VSPACESECTION}
\subsection{Investigated Object Detection R-CNNs}

\begin{figure}[t]
	\centering
	\begingroup
	\Large
	\resizebox{0.48\textwidth}{!}{
		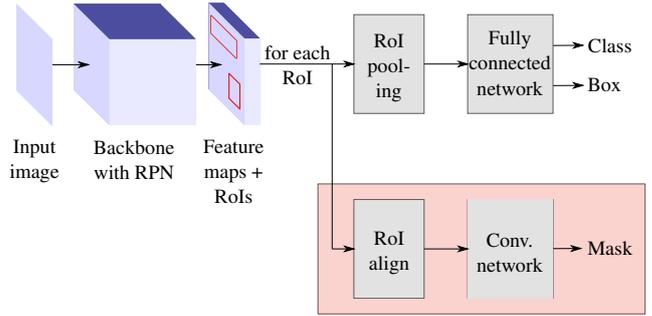
	}
	\endgroup
	\caption{Structure of used R-CNNs; Upper branch is used for Faster R-CNN, while the lower red branch is additionally used for Mask R-CNN to gain the pixel accurate masks.}
	\label{fig:r-cnns}
	\vspace{\VSPACE}
	\vspace{-2mm}
\end{figure}

\def\subwidth{0.33}
\begin{figure*}[t]
	\centering
	\begin{subfigure}[t]{\subwidth\textwidth}
	\centering
	\includegraphics[width=\textwidth]{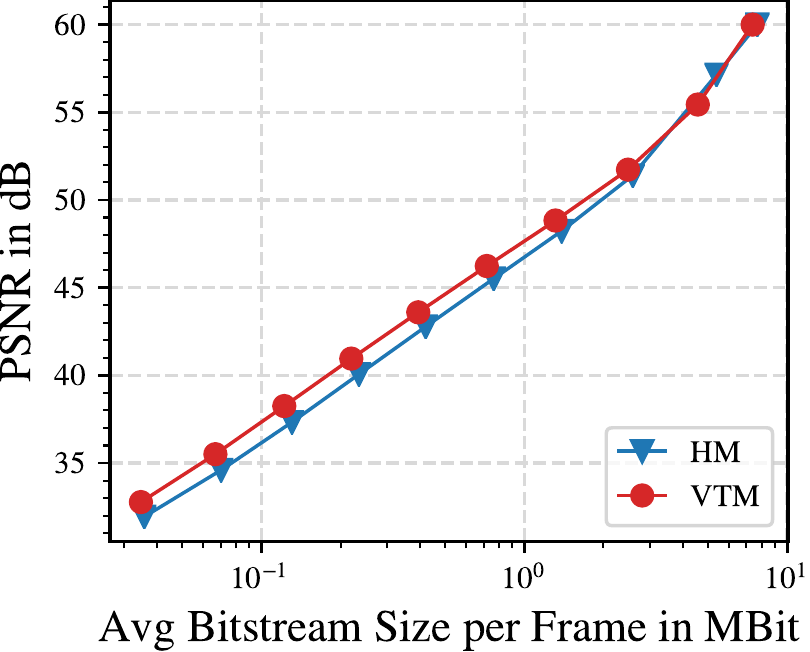}
	\end{subfigure}
	\begin{subfigure}[t]{\subwidth\textwidth}
	\centering
	\includegraphics[width=\textwidth]{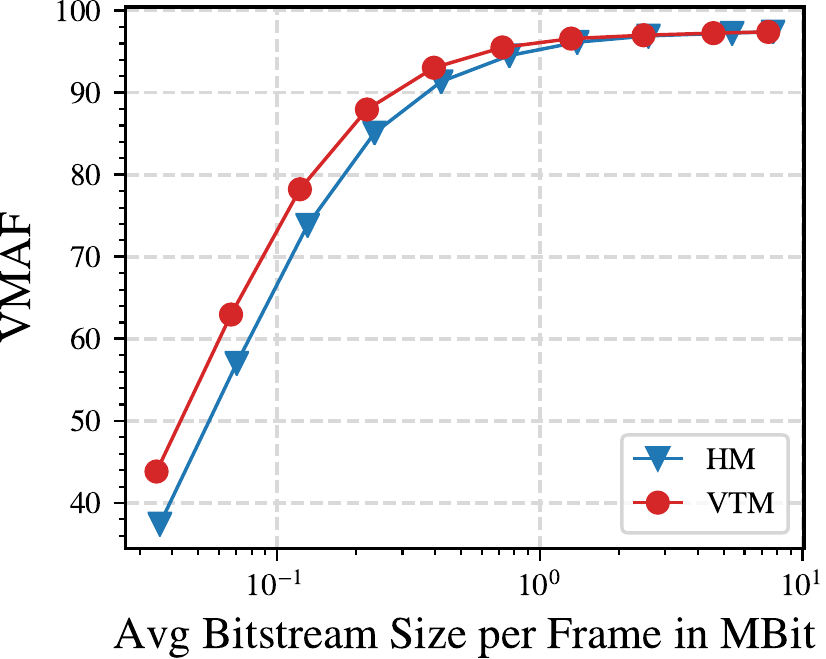}
	\end{subfigure}
	\begin{subfigure}[t]{\subwidth\textwidth}
	\centering
	\includegraphics[width=\textwidth]{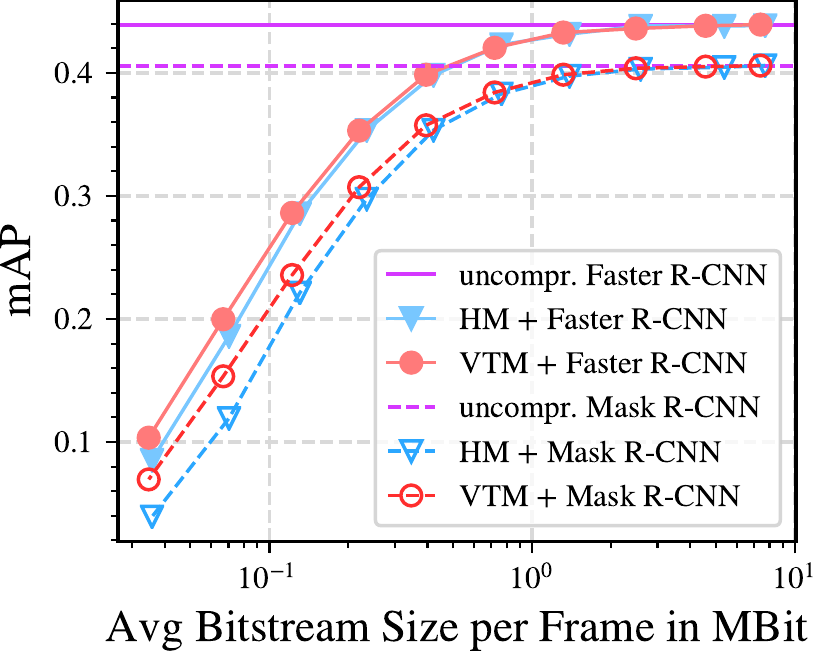}
	\end{subfigure}
	\caption{Rate-distortion diagrams with respect to PSNR, VMAF, and mAP comparing HM 16.2 with VTM 6.0 for 500 \textit{Cityscapes} validation images and all intra configuration for $\QP\in\{2, 7, 12, 17, 22, 27, 32, 37, 42, 47\}$.}
	\label{fig:res}
	\vspace{\VSPACE}
\end{figure*}

The state-of-the-art neural object detection networks Faster R-CNN \cite{ren2017} and Mask R-CNN \cite{he2017} were chosen to evaluate the influence of intra-frame compression on object detection.

Faster R-CNN outputs a class label with a certainty score and a bounding box for each potential object as depicted in the upper branch of Fig. \ref{fig:r-cnns}. To that extent, a feature map is extracted from the input image by a backbone consisting of convolutional layers that are adapted from classification networks. From this features the region proposal network~(RPN) derives regions of interest (RoIs), which are then classified and refined by the subsequent fully connected layers.



Mask R-CNN is a derivative of the Faster R-CNN structure targeting the problem of semantic segmentation with per-pixel classification. To create the pixel-accurate masks, a parallel convolutional network is attached to the Faster R-CNN structure as shown in Fig.~\ref{fig:r-cnns}. More detailed information on both used networks can be found in \cite{ren2017} and \cite{he2017}.


We use the \textit{PyTorch} implementations from the \textit{Detectron2} library \cite{wu2019detectron2} for both R-CNNs. For Mask R-CNN, we take the already existing model that has been trained on the \textit{Cityscapes} training set. Its backbone is a Residual Net  with 50 layers~(ResNet-50) \cite{he2016resnet} and Feature Pyramid Network~(FPN) \cite{lin2017fpn} structure. The same backbone is taken for Faster R-CNN, but since there is no pre-trained \textit{Cityscapes} model available in~\cite{wu2019detectron2}, the already existing model trained on the COCO dataset is taken from~\cite{wu2019detectron2} as initial weights. Subsequently, this model is further trained on the \textit{Cityscapes} classes and training images, where the GT pixel masks have been converted to bounding boxes before, for 42000 iterations, a batch size of 7 images, and a learning rate of $0.00025$.
\vspace{-4mm}
\vspace{\VSPACESECTION}
\section{Experimental Results}

\vspace{\VSPACESECTION}
\vspace{0mm}
\subsection{Comparison HEVC vs. VVC}


\begin{figure*}[t]
	\centering

	\includegraphics[width=0.8\textwidth]{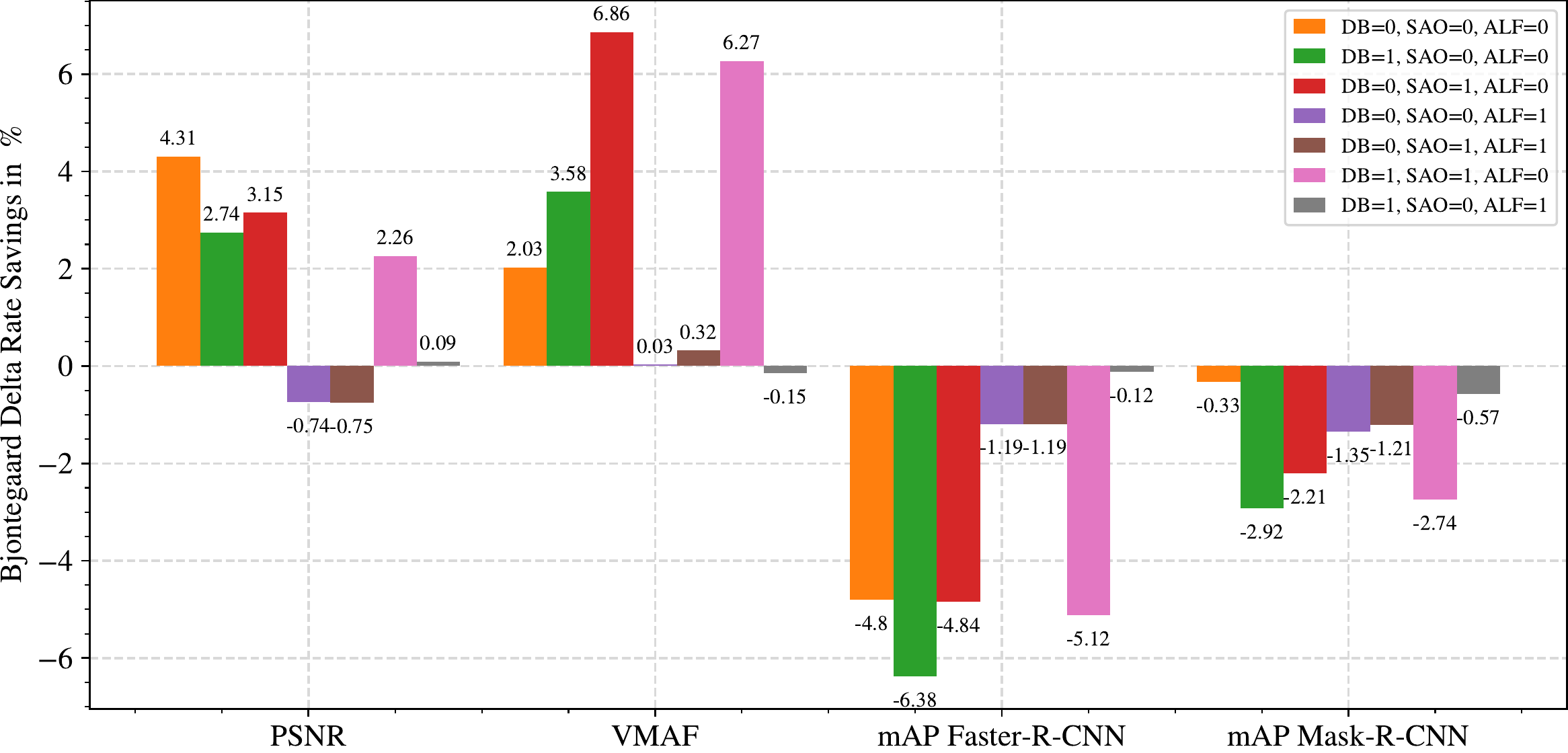}
	
	\caption{BDR in \% with respect to PSNR, VMAF, and mAP for Faster and Mask R-CNN comparing all possible permutations of activated or deactivated in-loop filters using the standard VTM with all in-loop filters activated as anchor; Used $\QP\in\{22, 27, 32, 37\}$; Negative BDR values represent bitrate savings compared to the anchor.}
	\label{fig:bd rate loop filters}
	\vspace{\VSPACE}
\end{figure*}
The performance of HM and VTM are compared by plotting peak signal-to-noise ratio (PSNR), Video Multi-Method Assessment Fusion (VMAF) with the \textit{vmaf\_v0.6.1.pkl} model~\cite{vmaf_git} and mAP over the bitrate. These rate-distortion curves for $\QP$ values from 2 to 47 in steps of 5 can be found in Fig.~\ref{fig:res}.


For object detection, the mAP does not significantly decrease for input images compressed with a $\QP$ lower than 17 compared to the uncompressed input data. With higher $\QP$ than 17, the network precision is dropping constantly, where VMAF shows a similar course. Contrary, PSNR continues to increase with higher bitrates.

In Table \ref{tab: BDR QP}, Bj\o ntegaard Delta Rate (BDR) \cite{bjontegaard2001} values for $\QP\in\{22, 27, 32, 37\}$ are listed as suggested by JVET \cite{bossen2019}. Higher $\QP$ are not considered, since they result in mAP values that are too small for practical applications. The BDR compares two rate-distortion curves by calculating the average bitrate savings for the same quality between two codecs. Originally, PSNR is taken as quality metric for calculating the BDR, however, we also use VMAF, mAP, and $\mathrm{mAP}_{50}$ analogously as it has been done in~\cite{choi2018}. Considering these BDR values, the coding gains between VTM and HM based on VMAF and PSNR above 22~\% are superior to taking the mAP into account (up to 13.6~\%). Hence, the optimizations made for the VTM are considerably more effective for the HVS than for neural detection networks, because the VTM is mainly optimized for delivering a decent PSNR value which does not necessarily results in a high detection rate. Since the coding artifacts have a higher influence on the task of pixel-wise segmentation than for drawing bounding boxes around the objects, the BDR savings for Mask R-CNN are higher than for Faster R-CNN. The coding gains for mAP are higher compared to $\mathrm{mAP}_{50}$, since the mask or the bounding box has to fit more precisely for mAP and thus coding artifacts are more disturbing for object detection.

\begin{table}[]
	\vspace{0.5mm}
	\caption{BDR in \% with respect to the particular quality metric using HM with the corresponding R-CNN as anchor for $\QP\in\{22, 27, 32, 37\}$.}
	\label{tab: BDR QP}
	\centering
	\begin{tabular}{l|llll}
		\hline
		& PSNR                    & VMAF                    & $mAP$    & $mAP_{50}$ \\ \hline
		Faster R-CNN          & \multirow{2}{*}{-22.17} & \multirow{2}{*}{-25.55} & -6.01  & -5.79   \\
		Mask R-CNN          &                         &                         & -13.56 & -11.24  \\ \hline
	\end{tabular}
	\vspace{\VSPACE}
	\vspace{-1mm}
\end{table}

\vspace{\VSPACESECTION}
\vspace{-1mm}
\subsection{Influence of In-Loop Filters on R-CNNs}

Derived from the previous results, additional investigations are taken whether existing VVC tools are actually improving the mAP when coding for R-CNNs. To that extent, the influence of the three VVC in-loop filters on object detection is compared with their influence on the HVS. The three in-loop filters are the de-blocking filter (DB) \cite{norkin2012} minimizing blocking artifacts, sample adaptive offset filter (SAO) \cite{fu2012} categorizing pixels and additionally transmitting suitable offset values, and the adaptive loop filter (ALF) \cite{tsai2013} convolving the output with a suitable filter as the last step of the coding chain. 

In Fig. \ref{fig:bd rate loop filters}, the resulting BDR savings for each permutation of active in-loop filters for $\QP\in\{22, 27, 32, 37\}$ are shown, taking the standard VTM with all in-loop filters activated as anchor. For BDR with respect to PSNR, deactivating in-loop filters results in decreased rate-distortion performance with respect to standard VTM for most permutations. However, there are two permutations with activated ALF and deactivated DB filter (purple and brown) that show minor bitrate savings compared to the standard VTM. Considering VMAF, deactivating the in-loop filters also results in higher bitrates at same VMAF and in the worst case 6.86~\% more bitrate is required.

Taking the in-loop filter influence on BDR for Faster R-CNN into account, the opposite behavior can be observed. There, deactivating in-loop filters results in high bitrate savings at the same detection rate for Faster R-CNN. When only activating the DB filter (green), 6.38~\% bitrate can be saved compared to the standard VTM. When using Mask R-CNN to evaluate the codec performance, the results are similar to Faster R-CNN since all permutations with at least one in-loop filter deactivated require less bitrate than the standard VTM. Again, only activating the DB filter results in the highest bitrate savings of 2.92~\% at the same mAP.

These results indicate that the R-CNNs are mainly sensitive to blocking artifacts while other artifacts that are reduced by SAO and ALF do not decrease the detection rate significantly. Contrary to BDR with respect to PSNR or VMAF, additionally activating the SAO or the ALF to the DB filter has no positive effect on the detection quality and also requires extra bits that lowers the coding performance. The bitrate savings for Faster R-CNN are higher than for Mask R-CNN when deactivating SAO and ALF since the task of drawing bounding boxes is less prone to coding artifacts than pixel-wise segmentation and thus the in-loop filters have less influence. All in all, it can be recommended deactivating the SAO and ALF when coding data for neural detection networks.



\vspace{\VSPACESECTION}

\section{Conclusion}

Video coding for machines is a relevant topic, and rate-mAP curves for state-of-the-art codecs and R-CNNs were presented within this paper and compared against HVS metrics. The first experiment showed that the large BDR savings above 20~\% comparing VVC against HEVC can only be observed for the metrics that represent the human as final user. For the case that R-CNNs analyze the coded image, the coding gains were not that high, although they are still between 5~\% and 14~\% depending on the used network and mAP metric. Furthermore, it was found that when deactivating all in-loop filters except the DB filter, significant bitrate savings can be achieved when coding for R-CNNs. Additionally, computational complexity can be saved omitting the SAO and the ALF. All in all, new VVC optimization methods have to be found when coding for neural networks in order to achieve bitrate savings in the same dimension as for PSNR and VMAF. Possible approaches might consider replacing the PSNR in the rate-distortion optimization of the VVC with metrics better reflecting the R-CNN behavior. Besides, it should be investigated whether the presented results are also valid for inter coding. However, this requires a suitable labeled and uncompressed video dataset.

%


%

\clearpage
\bibliographystyle{IEEEbib}
\setstretch{0.5}
\bibliography{/home/fischer/Paper/jabref_literature_research_ms2.bib,/home/fischer/Paper/literature_M2M_communication.bib}

\end{document}